\documentclass[9pt,twocolumn,twoside]{pnas-new}

\templatetype{pnasresearcharticle} 

\begin{document}

\title{Multiflagellarity leads to the size-independent swimming speed of peritrichous bacteria}


\author[a,1,2]{Shashank Kamdar}
\author[a,1,2]{Dipanjan Ghosh}
\author[b]{Wanho Lee}
\author[c,2]{Maria T\u{a}tulea-Codrean}
\author[d]{Yongsam Kim}
\author[a]{Supriya Ghosh}
\author[a]{Youngjun Kim}
\author[a]{Tejesh Cheepuru}
\author[c]{Eric Lauga}
\author[e,2]{Sookkyung Lim}
\author[a,2]{Xiang Cheng}

\affil[a]{Department of Chemical Engineering and Materials Science, University of Minnesota, Minneapolis, MN 55455, USA}
\affil[b]{National Institute for Mathematical Sciences, Daejeon 34047, Republic of Korea}
\affil[c]{Department of Applied Mathematics and Theoretical Physics, University of Cambridge, Cambridge CB3 0WA, United Kingdom}
\affil[d]{Department of Mathematics, Chung-Ang University, Seoul 06974, Republic of Korea}
\affil[e]{Department of Mathematical Sciences, University of Cincinnati, Cincinnati, OH 45221, USA}

\leadauthor{Kamdar}

\significancestatement{A taller person usually swims faster thanks to the more pronounced increase of hydrodynamic thrust with height relative to fluid drag in inertia-dominated flows. However, this general positive size-speed correlation may not hold at microscopic scales for bacteria, where viscous dissipation overwhelms inertia in their swimming. By combining experiments, simulations and theory, we show that the swimming speed of bacteria with multiple flagella remains constant across the natural range of bacterial lengths, settling a long-lasting debate over the size-speed relation of bacterial swimming. Our quantitative analyses further elucidate the hydrodynamic origin of such an unexpected relation. Our study sheds light on the functional benefit of multiflagellarity and provides a useful guide for designing artificial microswimmers.}

\authorcontributions{S.K and X.C conceived the project. S.K., D.G. and X.C. designed the research and developed the models. S.K., D.G., S.G., Y.K., T.C. and X.C. performed experiments. W.L., Y.K. and S.L. performed the IB simulations. M.T.-C. and E.L. performed the SBT simulations. S.K., D.G. and X.C. co-wrote the manuscript with inputs from all the authors. S.L. and X.C. supervised the project.}
\authordeclaration{The authors declare no competing interests. }
\equalauthors{\textsuperscript{1}These authors contributed equally to this work.}
\correspondingauthor{\textsuperscript{2}To whom correspondence should be addressed. E-mail: kamda008@umn.edu, ghosh135@umn.edu, m.tatulea-codrean@damtp.cam.ac.uk, sookkyung.lim@uc.edu, xcheng@umn.edu}


\begin{abstract}
To swim through a viscous fluid, a flagellated bacterium must overcome the fluid drag on its body by rotating a flagellum or a bundle of multiple flagella. Because the drag increases with the size of bacteria, it is expected theoretically that the swimming speed of a bacterium inversely correlates with its body length. Nevertheless, despite extensive research, the fundamental size-speed relation of flagellated bacteria remains unclear with different experiments reporting conflicting results. Here, by critically reviewing the existing evidence and synergizing our own experiments of large sample sizes, hydrodynamic modeling and simulations, we demonstrate that the average swimming speed of \textit{Escherichia coli}, a premier model of peritrichous bacteria, is independent of their body length. Our quantitative analysis shows that such a counterintuitive relation is the consequence of the collective flagellar dynamics dictated by the linear correlation between the body length and the number of flagella of bacteria. Notably, our study reveals how bacteria utilize the increasing number of flagella to regulate the flagellar motor load. The collective load sharing among multiple flagella results in a lower load on each flagellar motor and therefore faster flagellar rotation, which compensates for the higher fluid drag on the longer bodies of bacteria. Without this balancing mechanism, the swimming speed of monotrichous bacteria generically decreases with increasing body length, a feature limiting the size variation of the bacteria. Altogether, our study resolves a long-standing controversy over the size-speed relation of flagellated bacteria and provides new insights into the functional benefit of multiflagellarity in bacteria. 
\end{abstract}

\dates{This manuscript was compiled on \today}
\doi{\url{www.pnas.org/cgi/doi/10.1073/pnas.XXXXXXXXXX}}

\maketitle
\thispagestyle{firststyle}
\ifthenelse{\boolean{shortarticle}}{\ifthenelse{\boolean{singlecolumn}}{\abscontentformatted}{\abscontent}}{}

\firstpage[11]{2}

\dropcap{T}he motility of bacteria is an essential part of their strategy for survival and competition in dynamic environments \cite{Bray2000,Hibbing2010,Keegstra2022}. This locomotion at microscopic scales is directly relevant to macroscopic ecological and biomedical processes such as bio-remediation, oceanic carbon cycle and spread of infections, which are of great consequence to human well-being \cite{Grossart2001,Ottemann2002,Josenhans2002,Lemon2007}. As a paradigmatic model for bacterial motility, flagellated bacteria swim by rotating a helical rod-shaped propeller composed of either a single flagellum (monotrichous bacteria) or a bundle of multiple flagella (peritrichous bacteria) \cite{Berg2004}. Driven by individual motors, these flagella collectively rotate to generate hydrodynamic thrust, which balances the fluid drag experienced by bacteria and enables their swimming \cite{Lauga2020}. At the same time, the torque driving the flagellar rotation is offset by the torque driving the rotation of the body to ensure zero net torque on a bacterium. The flagellar motors carry these torques as a load following a characteristic torque-speed relation \cite{Berry2008}. Although it is generally expected that the size of flagellated bacteria affects their swimming speed due to the dependence of fluid drag on body shape dictated by universal hydrodynamic principles \cite{Lauga2020}, the relation between the body size of a bacterium and its swimming speed is still under debate. A consensus on even the {\it qualitative} trend of the relation has not yet been reached, with studies reporting positive \cite{Leonardo2011}, negative \cite{Guadayol2017} and no correlation \cite{Kaya2012} between size and speed covering all the possible monotonic trends.

It is surprising that a controversy over such a fundamental relation of the kinematics of bacterial motility persists following the decades of research on the hydrodynamics of swimming flagellated bacteria (see \cite{Lauga2020} and references therein), especially considering the implication of the size-dependent swimming speed on the evolution and ecology of bacteria \cite{Keegstra2022}. As one of the key phenotypes of bacteria, bacterial body size shows natural variations even in a genetically identical population growing in a constant homogeneous environment \cite{Jun2018}. These variations arise from the stochastic nature of gene expression and the fluctuation of generation times. Since the chemotactic ability of a bacterium is positively correlated with its swimming speed \cite{Lovely1975,Celani2010,Son2016}, a size-dependent swimming speed would indicate a size-dependent chemotactic ability. Hence, the natural size variation of a bacterial population would result in the diversity of chemotactic behaviors in the population. Such a phenotypic diversity is known to strongly affect the overall fitness of the population for selection and needs to be closely regulated in ecological settings \cite{Fu2018,Waite2018}. 

\begin{SCfigure*}[\sidecaptionrelwidth][t!]
\centering
\includegraphics[width=11.4cm,height=6.3cm]{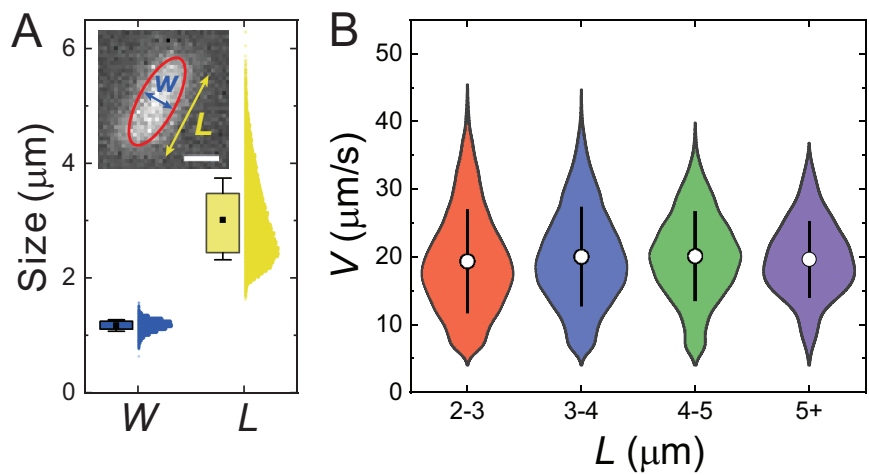}
\caption{Size-independent swimming speed. (\textit{A}) The distributions of the width $W$ and the length $L$ of the cell bodies of \textit{E. coli} from a genetically identical colony. Inset: A micrograph of a fluorescently labeled bacterium with $W$ and $L$ indicated. The scale bar is 1 $\mu$m. (\textit{B}) A violin plot showing the distributions of the swimming speed $V$ of \textit{E.~coli} of different lengths. The population average swimming speeds (the white dots) and the standard deviations of the distributions (the black lines) are independent of $L$.}\label{fig:fig1}
\end{SCfigure*}

Alongside body size, the number and the morphology of bacterial flagella also profoundly influence the swimming speed of bacteria \cite{Darnton2007}. Thus, potential correlations between body size and flagellar architecture substantially complicate the understanding of the size-speed relation and pose a major challenge for resolving the controversy. Studies that independently investigated the dependence of the swimming speed on the size of bacteria \cite{Leonardo2011,Guadayol2017,Kaya2012} or the architecture of flagella \cite{Reigh2012,Kanehl2014,Nguyen2018} failed to deliver a coherent picture of the swimming behaviors of flagellated bacteria. Recognizing the deficiency of the previous approach, here we combine existing experimental evidence with our own high-throughput experiments, hydrodynamic models and simulations and conduct a comprehensive study of the interrelation between physical features (the body size and the architecture of flagella) and swimming characteristics (the swimming speed and the angular speeds of body and flagella) of \textit{Escherichia coli} (\textit{E. coli}), the most studied model of peritrichous bacteria for bacterial motility. Our study reveals that the linear correlation between the size of bacteria and the number of flagella enforces load sharing among different flagellar motors, which regulates the collective dynamics of multiple flagella and sets the size-independent swimming speed of bacteria. As such, our work provides a unified resolution to the long-standing controversy and sheds light on the hydrodynamic origin of the size-speed relation of flagellated bacteria.

Our paper is organized as follows. First, we present our experimental findings on how the swimming speed of bacteria depends on their body length and discuss our results in the context of the existing literature. We then illustrate the origin of the size-speed relation via two sets of low-Reynolds-number hydrodynamic models and simulations. Finally, we consider the biological implications of our results on the swimming efficiency of bacteria, the selective advantage of multiflagellarity, and differences between peritrichous and monotrichous bacteria in swimming and morphology.

\section*{Results}

\subsection*{Size-speed relation of flagellated bacteria} To resolve the controversy over the relation between body size and swimming speed of flagellated bacteria, we image the locomotion of a large number of \textit{E. coli} cells ($> 25,000$) in bulk fluids away from system boundaries and measure the swimming speed of bacteria, $V$, in the run phase of their classic run-and-tumble motion (Materials and Methods) \cite{Berg2004}. The body length, $L$, of \textit{E. coli} displays a broad right-skewed distribution with a median of 2.87 $\mu$m (Fig.~\ref{fig:fig1}\textit{A}), which allows us to probe the correlation between $V$ and $L$ over threefold bacterial size. In comparison, the distribution of body width is much narrower, which we treat as a constant. At a given body length, the swimming speed of individual bacteria varies strongly (Fig.~\ref{fig:fig1}\textit{B}), owing to the cell-to-cell variations in metabolism and the number and morphology of flagella \cite{Kamdar2022_thesis}. Thanks to the large sample size of our experiments, we are able to measure the population-average swimming speeds and the speed distributions accurately despite this strong variation. Particularly, we find that both the average swimming speeds and the speed distributions are independent of $L$, exhibiting no correlation with the body length (Fig.~\ref{fig:fig1}\textit{B}). This size-independent swimming speed is quite counterintuitive: as the fluid drag coefficient on a bacterial body increases by more than $60\%$ when the body length increases from 2 to 6 $\mu$m (Eq. (9) in Supplementary Information (SI)), theories and simulations of bacterial swimming would predict more than $30\%$ decrease in $V$ when the architecture of the flagellar bundle is fixed (Fig.~S1).

Using our data-intensive results as a benchmark, we revisit previous studies that explored the size-speed relation of \textit{E. coli}. Di Leonardo and co-workers first measured the swimming speed of \textit{E. coli} as a function of the body length of bacteria near a fluid-air interface \cite{Leonardo2011}. They reported a weak increase of $V$ with $L$. Nevertheless, the sample size of their measurements was small ($\sim 50$ bacteria in total), which explains the large uncertainties of their measurements that may obscure the genuine trend of $V(L)$. Kaya and Koser reported the swimming speed of \textit{E. coli} near the solid boundaries of microfluidic channels in external shear flow \cite{Kaya2012}. They showed that $V$ is independent of $L$, agreeing with our measurements. However, it is not clear how solid boundaries and shear flow in their experiments affect the swimming behaviors of bacteria \cite{Lauga2006,Marcos2012,Rusconi2014}. More recently, Guadayol \textit{et~al.} measured the size-dependent swimming speed of both wild-type and filamentous \textit{E. coli} treated with antibiotics in bulk fluids \cite{Guadayol2017}. The authors reached a conclusion that $V$ decreases with $L$ based on their results of filamentous bacteria with $L$ up to 38 $\mu$m, although their measurements on wild-type bacteria do show a constant swimming speed independent of $L$. For antibiotic-treated filamentous bacteria, the body length of bacteria is larger than the length of flagella ($\sim 6$ to 8 $\mu$m). Flagella distributed randomly along the entire body surface cannot form a single bundle \cite{Maki2000}. Hence, the swimming mechanism of filamentous bacteria is qualitatively different from that of natural-sized wild-type bacteria \cite{Phan2018}. In this study, we focus on bacteria of natural sizes, whose swimming is driven by the rotation of a single bundle. 

To solve the puzzle of the size-independent swimming speed, we explore the key role of multiflagellarity in bacterial swimming next.

\subsection*{Correlation between flagellar number and bacterial length}

\begin{figure}
\centering
\includegraphics[width=0.9\linewidth]{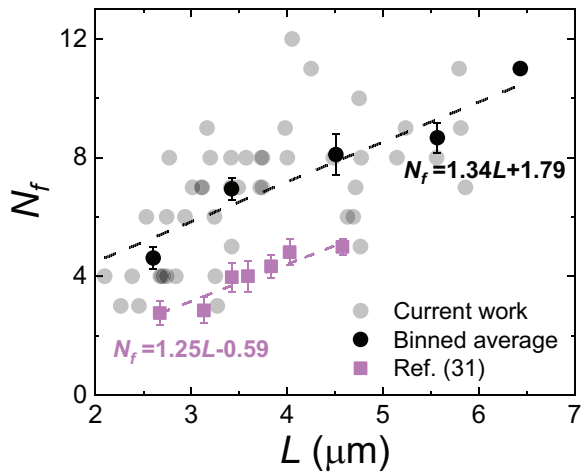}
\caption{Multiflagellarity. The number of flagella $N_f$ versus the body length $L$ of {\it E. coli}. Gray disks are individual measurements, whereas black disks and purple squares are the binned average values. The dashed lines are the linear regressions of the binned data.}
\label{fig:fig2}
\end{figure}

We measure the number of flagella, $N_f$, for bacteria of different $L$ using transmission electron microscopy (TEM) (Materials and Methods), which shows a positive linear correlation (Fig.~\ref{fig:fig2}),
\begin{equation}
   N_f = 1.34L + 1.79. \label{eq:eqNf}
\end{equation} 
Here, $L$ is measured in units of microns. This positive correlation provides the first clue towards the origin of size-independent swimming speed. 

The linear relation of Eq.~\ref{eq:eqNf} is consistent with the previous finding that flagella are randomly distributed along the length of the bacterial body \cite{Maki2000}. A similar linear relation has also been observed by Mears (Fig.~\ref{fig:fig2}) \cite{Mears2014_thesis}. However, a recent study found that bacteria under different growth conditions show varying average body size but the same average number of flagella \cite{Honda2022}. This discrepancy suggests that the finding of the study may not be applicable to the cell-to-cell variation in a single population under the same growth condition studied here in our experiments.

\subsection*{Collective flagellar dynamics}

\begin{figure*}
\centering
\includegraphics[width=\linewidth]{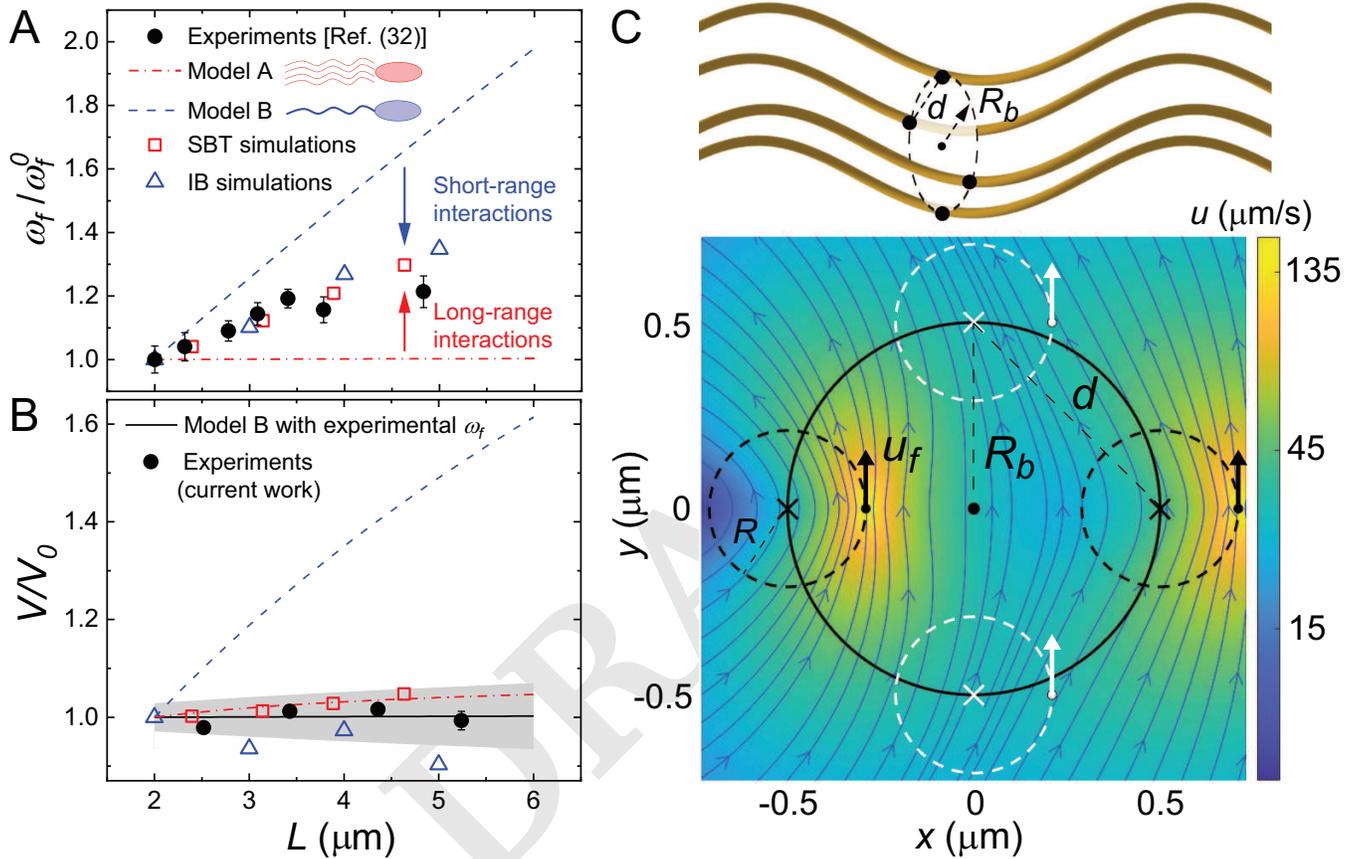}
\caption{Collective dynamics of multiple flagella in a bundle. ({\it A}) Angular speed of flagella $\omega_f$ as a function of body length $L$ of bacteria. $\omega_f$ is normalized by $\omega_f$ at $L = 2$ $\mu$m, $\omega_f^0$. Empty symbols are for simulations, and solid disks for experiments extracted from Ref.~\cite{Chattopadhyay2006}. Model A: an extended Purcell model with independently rotating flagella. Model B: the classic Purcell model with a single rotating bundle. ({\it B}) Swimming speed $V$ as a function of $L$. $V$ from the models is normalized by $V$ at $L = 2$ $\mu$m, $V_0$, whereas experimental $V$ extracted from Fig.~\ref{fig:fig1}{\it B} is normalized by the average $V$ over all $L$. The lines and symbols are the same as those used in ({\it A}). The solid line is the prediction of Model B using the experimental $\omega_f$ as input. The shaded region indicates the uncertainty due to the experimental errors of $\omega_f$. ({\it C}) Arrangement of flagella in the SBT simulations and mechanism for hydrodynamic enhancement of rotation. Top: A schematic of the flagellar arrangement (not to scale). The axes of the helical flagella are normal to the plane ($x$-$y$) and located evenly around a circle of radius $R_b$ with an inter-flagellar distance $d = 2R_b\sin(\pi/N_f)$. The number of flagella $N_f = 4$ in the schematic. The simulation results in ({\it A}) and ({\it B}) have $R_b = 0.5$ $\mu$m and $N_f(L)$ from Eq.~\ref{eq:eqNf}. Bottom: Two fictitious flagella (white dashed circles) are hydrodynamically entrained into a higher rotation speed due to the flows induced by two actual flagella (black dashed circles). The background color and the streamlines indicate, respectively, the magnitude and direction of the flow velocity $u$ induced by the actual flagella in the $x$-$y$ plane. Note that the thick arrows, which represent the in-plane velocity of the flagella along circular trajectories (dashed circles), $u_f = \omega_fR$, form acute angles with the direction of local flow (blue streamlines). At fixed motor torque, this results in the hydrodynamic enhancement of rotation for the fictitious flagella. Here, $R$ is the helical amplitude of flagellar filaments.}
\label{fig:fig3}
\end{figure*}

Deciphering the origin of the size-independent swimming speed requires a quantitative understanding of the collective dynamics of multiple flagella in a bundle. Chattopadhyay and co-workers measured the angular speed of the flagellar bundle $\omega_f$ as a function of the body length of {\it E. coli} using an optical trap \cite{Chattopadhyay2006}, where they found $\omega_f$ increases with $L$ (Fig.~\ref{fig:fig3}{\it A}). Such an increasing trend needs to be combined with the linear correlation between $N_f$ and $L$ to form a coherent picture of the size-speed relation of flagellated bacteria.  

The motion of individual flagella in a bundle must be coordinated to ensure the smooth rotation of the bundle as a whole. In addition to the geometric constraints imposed by the helical shape \cite{Macnab1977}, the inter-flagellar interactions at the short distance of tens of nanometers in a bundle involve complex hydrodynamic, elasto-hydrodynamic and steric interactions \cite{Berg1973,Kim2003,Reigh2012,Ishimoto2019,Tatulea-Codrean_2022}, which cannot be solved analytically. To overcome the difficulty, we begin by examining two limits that allow for exact solutions: Model A considers independently rotating non-interacting flagella, whereas Model B deals with a single rotating bundle without relative inter-flagellar motions. The two models reveal two opposing factors respectively, the competition of which determines the collective dynamics of multiple flagella in a rotating bundle. To further understand such a competition, we conduct two different types of low-Reynolds-number hydrodynamic simulations, one following each of the two analytical models. Starting from the opposite limits, the two simulations converge and quantitatively match the experimental findings.

{\it Model A}: We first consider $N_f$ non-interacting flagella that rotate independently (SI Sec.~1A). We extend the Purcell model of flagellated bacteria and use resistive force theory (RFT) to calculate the drag coefficients of flagellar filaments \cite{Purcell1997}. In Model A, bacterial swimming speed increases weakly with $L$, exhibiting an approximately size-independent swimming speed (Fig.~\ref{fig:fig3}{\it B}). However, the angular speed of flagella $\omega_f$ is also independent of $L$, contrary to the increasing trend of $\omega_f(L)$ observed in experiments (Fig.~\ref{fig:fig3}{\it A}) \cite{Chattopadhyay2006}. 

The increase of $\omega_f(L)$ is caused by the fluid-mediated inter-flagellar interactions ignored in Model A, an effect we illustrate via simulations of interacting flagella at different inter-flagellar distances (SI Sec.~2A) \cite{Tatulea-Codrean2021}. Specifically, we prescribe the arrangement of flagella by placing $N_f$ rigid rotating helical filaments evenly around a circle of control radius $R_b$ with the inter-flagellar distance $d = 2R_b\sin(\pi/N_f)$ (Fig.~\ref{fig:fig3}{\it C}). The flagellar assembly and the body are coupled through the force and torque balance, whereas the drag coefficients of the flagellar assembly are calculated numerically using slender-body theory (SBT). Figure~\ref{fig:fig3}{\it C} shows the flow field induced by two rotating flagella in the plane ($x$-$y$) normal to the axes of the flagellar helices ($z$). The trajectories of the rotating flagella within the plane are circular as indicated by the black dashed circles. If two more flagella were added to the assembly (the white dashed circles in Fig.~\ref{fig:fig3}{\it C}), they would experience less viscous resistance from the already entrained fluid, because the projection of the velocity of the original flow to the moving direction of the new flagella is positive at the positions of the new flagella. Thus, coupled by hydrodynamic interactions, the flagella rotate faster together than they would on their own assuming a constant motor torque. 

Quantitatively, we confirm that $\omega_f$ increases with $N_f$, which in turn increases with $L$ due to the linear correlation between $N_f$ and $L$ (Fig.~\ref{fig:fig3}{\it A}). Such an increasing trend becomes more obvious as the hydrodynamic coupling grows stronger with reducing $d$ (Fig.~S3{\it A}). Our SBT simulations thus demonstrate that the long-range hydrodynamic coupling between flagella enhances flagellar rotation. As we ignore the torque density along the filament centerline that would contribute a higher-order correction to the flow, the simulations cannot capture the effect of viscous dissipation induced by the relative rotation of flagella at short distances \cite{Purcell1997,Berg1973}. Hence, at small $d$ comparable to the inter-flagellar distance of real bacterial bundles, the SBT simulations overpredict the trend of $\omega_f(L)$ (Fig.~S3{\it A}).

{\it Model B}: To describe flagellar dynamics at small $d$, we apply the classic Purcell model in the second limit, where $N_f$ flagella combine into a tight bundle without relative motions (SI Sec.~1B). In Model B, $\omega_f$ increases substantially with $N_f$ (Fig.~\ref{fig:fig3}{\it A}), which sets the upper bound on the flagellar angular speed of the SBT simulations as $d \to 0$ (Fig.~S3{\it A}). The strong increase of $\omega_f$ also leads to a strong increase of $V$ with $L$ (Fig.~\ref{fig:fig3}{\it B}). Interestingly, if we insert the experimental $\omega_f(L)$ as an input into the interrelations $V(\omega_f)$ and $\omega_b(\omega_f)$ predicted by Model B (Eqs. 14 and 15 in SI), we recover not only the size-independent swimming speed (Fig.~\ref{fig:fig3}{\it B}) but also the dependence of the angular speed of bacterial body on the body length, $\omega_b(L)$ (Fig.~S2). 

To account for the short-range viscous and steric interactions induced by the relative motions of flagella in a bundle \cite{Macnab1977}, we conduct a second set of simulations in the framework of the immersed boundary (IB) method to examine the formation of a flagellar bundle from $N_f$ initially-separated elastic helical filaments (SI Sec.~2B, Supplementary Videos 1 and 2) \cite{Lee2021}. In our IB simulations, each of the filaments is connected normal to the surface of a spherocylindrical cell body via an elastic hook (Fig.~S4). At each $L$, we match the initial angular speed of separated filaments with $\omega_f(L)$ calculated from Model B by adjusting the flagellar motor torque \cite{Lee2021}. This initial condition imitates the rotation of filaments in an ideal bundle without short-range inter-flagellar interactions. We then obtain $\omega_f(L)$ after the formation of the stable flagellar bundle when the short-range interactions fully set in. Contrary to the rotation-enhancing effect of the long-range hydrodynamic coupling, the short-range viscous and steric interactions reduce $\omega_f(L)$ relative to the prediction of Model B, leading to a balanced result that quantitatively agrees with experimental measurements (Fig.~\ref{fig:fig3}{\it A}) \cite{Chattopadhyay2006}. 

Finally, both the SBT simulations at a chosen $R_b$ and the IB simulations show an approximately size-independent swimming speed, agreeing with our experiments (Fig.~\ref{fig:fig3}{\it B}).   

\section*{Discussion}

\subsection*{Motor performance}

\begin{figure}
\centering
\includegraphics[width=1\linewidth]{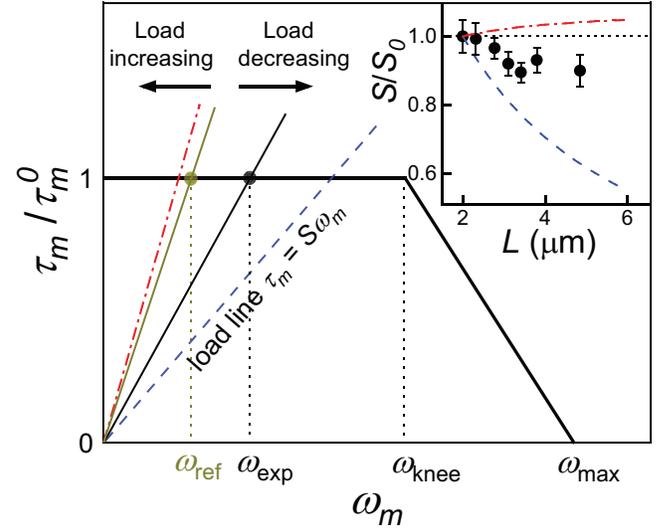}
\caption{Flagellar motor performance. A schematic of the flagellar motor performance curve connecting the motor torque $\tau_m$ and speed $\omega_m$. $\tau_m$ reaches an approximate plateau $\tau_m^0$ at low $\omega_m$. The knee speed $\omega_{knee}$ and the maximum speed $\omega_{max}$ are indicated. Intersection of the performance curve with the load line $\tau_m = S\omega_m$ determines the motor operating speed. A reference load line at bacterial body length $L = 2$ $\mu$m is indicated by the brown solid line. In comparison, three load lines at $L = 5$ $\mu$m are shown, which represent the experiments and the two model calculations in Fig.~\ref{fig:fig3} respectively following the same line styles. Inset: the slope of the load line $S$ as a function of $L$ from the experiments and the two models. $S$ is normalized by the slope at $L = 2$ $\mu$m, $S_0$. The horizontal dotted line indicates a size-independent load $S/S_0 = 1$.}
\label{fig:fig4}
\end{figure}

The complex flagellar dynamics discussed above can be physically understood from the motor performance curve characterizing the torque-speed relationship of individual flagellar motors (Fig.~\ref{fig:fig4}) \cite{Berg2004,Berry2008}. The torque output of a flagellar motor $\tau_m$ is approximately constant with increasing motor angular speed $\omega_m = |\omega_f – \omega_b|$ up to a knee speed $\omega_{knee} \approx 1300$ rad/s at room temperature, beyond which $\tau_m$ decreases linearly and reaches zero at $\omega_{max}$. Flagellar motors operate below $\omega_{knee}$ with a constant torque $\tau_m^0$ for swimming bacteria in our study. The drag torque on the body, together with the torque balance between the body and the bundle, specifies a linear load line with its slope depending on $N_f$, the fluid viscosity and the geometry of the body and the bundle (SI Sec.~1) \cite{TatuleaCodrean2021_thesis}. The intercept between the load line and the motor performance curve determines the operating $\omega_m$ of flagella. Note that, since the rotational drag coefficient of the body is substantially larger than that of the flagella, $\omega_b \ll \omega_f$. Thus, $\omega_f$ follows closely the trend of $\omega_m$.

The different scenarios of flagellar configurations in the Results section can now be reconsidered in terms of motor performance. First, for fixed $N_f$, the total load on the $N_f$ motors would increase with the body length due to the higher drag on the body, which would raise the slope of the load line on each motor and thus slow down $\omega_m$. However, with increasing $N_f(L)$ (Eq.~\ref{eq:eqNf}), the increased load is carried by more flagella. If all the flagella rotated independently without forming a bundle (Model A), the linear increase of $N_f(L)$ would approximately cancel the proportionally increased load of the body, resulting in an incorrect prediction of nearly constant load on each motor and therefore constant $\omega_m$ and $\omega_f$ (the red dashed-dotted line in Fig.~\ref{fig:fig4}, see Eq.~8 in SI for the calculation). In contrast, flagellar motors in a tight bundle (Model B) collectively share the increased load of the body (Eq.~16 in SI), thereby reducing the load on each motor. As a result, $\omega_m$ and $\omega_f$ increase with $L$. Since the viscous dissipation between flagella at short distances further raises the load carried by each flagellum, the load line of a tight bundle without short-range inter-flagellar interactions (Model B, the blue dashed line in Fig.~\ref{fig:fig4}) overtakes the experimental load line with such interactions (the black line in Fig.~\ref{fig:fig4}). Nevertheless, the hallmark of the collective load sharing between multiple flagella---the increasing trend of $\omega_f(L)$---is well preserved experimentally (Fig.~\ref{fig:fig3}{\it A}). Thus, by bringing the flagella closely together in an interacting bundle, bacteria effectively regulate the load shared by each flagellum in the bundle, which leads to faster flagellar rotation counterbalancing the higher rotational fluid drag on larger cell bodies and sustaining a constant swimming speed.    

\subsection*{Other possible mechanisms}

We rule out two alternative mechanisms as the possible explanation of the size-independent swimming speed. First, the decrease of wobbling angle of bacteria results in the increase of their swimming speed \cite{Kamdar2022}. Thus, a negative correlation between the wobbling angle and the body length of bacteria could give rise to the size-independent swimming speed of bacteria. We measure the wobbling angle of bacteria of different lengths (Fig.~S5) and show that the decrease of wobbling angle is not sufficient to explain the size-independent swimming speed (SI Sec.~3A).  

Second, it has been suggested previously that the change of the morphology of flagellar filaments with the length of bacteria may be the origin of the size-independent swimming speed \cite{Kaya2012,Turner2000}. We examine the morphology of flagellar filaments via experiments using fluorescently-labeled bacteria (Materials and Methods) and simulations (Fig.~S6). Both experimental and numerical results show that the change of the morphology is too small to explain the size-independent swimming speed (SI Sec.~3B).

\begin{figure}
\centering
\includegraphics[width=0.9\linewidth]{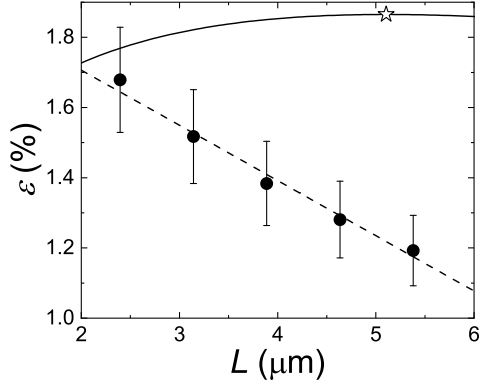}
\caption{Propulsion efficiency. Propulsion efficiency $\epsilon$ versus bacterial body length $L$. The dashed line is the linear regression of the data. The error bars represent standard errors. The solid line is the prediction of the classic Purcell model (Model B) with a constant number of flagella $N_f = 4$. A maximum efficiency of $\epsilon = 1.87\%$ is obtained when $L = 5.1$ $\mu$m (the star).}
\label{fig:fig5}
\end{figure}

\subsection*{Propulsion efficiency}

The efficiency of propulsion of microorganisms has been studied extensively because of its role in the optimization of the swimming kinematics and morphology of microorganisms \cite{Childress1981,Purcell1997,Chattopadhyay2006,Spagnolie2011,Tam2011}. It is defined as the ratio of the rate of work required to drag a swimmer at given velocity $V$ to the total mechanical power exerted by the swimmer on the fluid when swimming at the same speed \cite{Childress1981}. Specifically, the propulsion efficiency of flagellated bacteria such as \textit{E. coli} can be calculated as \cite{Chattopadhyay2006}
\begin{equation}
    \epsilon = \frac{A_b V^2}{N_f \tau_m^0 \omega_m},
\end{equation}
where $A_b$ is the translational drag coefficient of an ellipsoidal cell body given by Eq.~9 in SI and $V$, $N_f$ and $\omega_m$ are all taken directly from experiments. In addition, we adopt the constant motor torque $\tau_m^0 \approx 400$ pN$\cdot$nm based on our RFT calculation using $\omega_f$ from experiments as input. 

Although the swimming speed of bacteria is independent of their size, the propulsion efficiency decreases linearly with $L$ (Fig.~\ref{fig:fig5}). As $A_b/N_f$, $V$ and $\tau_m^0$ are approximately size-independent, $\epsilon(L) \sim 1/\omega_m(L)$. Driven by the increase of $\omega_m$ with $L$ (Fig.~\ref{fig:fig3}{\it A}), $\epsilon$ decreases with increasing $L$. Our experimental finding contradicts the prediction of previous studies where the correlation between $N_f$ and $L$ was ignored. The original Purcell model with a constant $N_f$ shows a non-monotonic trend of $\epsilon(L)$ (the solid line in Fig.~\ref{fig:fig5}). The existence of a maximum $\epsilon$ has been used to rationalize the size ratio between bacterial body and flagella \cite{Purcell1997,Chattopadhyay2006}. The monotonic decrease of the propulsion efficiency observed in our experiments challenges this explanation. We find that bacteria of small sizes have the highest propulsion efficiency, which is close to the maximum efficiency of the Purcell model and approaches the upper bound for flagellated bacteria at $2-3\%$ \cite{Lauga2020}.       

\begin{figure}
\centering
\includegraphics[width=0.9\linewidth]{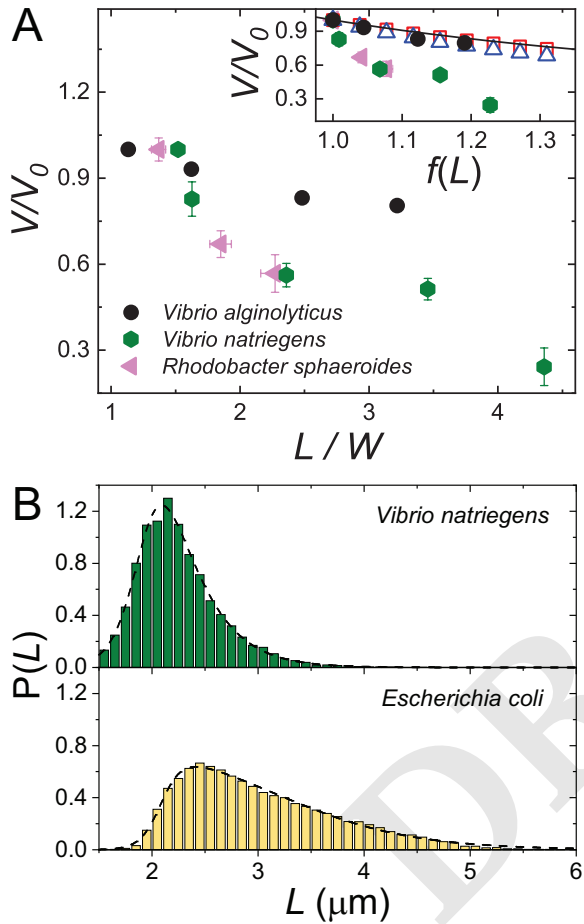}
\caption{Swimming speed and morphology of monotrichous bacteria. ({\it A}) The swimming speeds $V$ of three species of monotrichous bacteria as a function of their body length $L$. $V$ is normalized by the swimming speed at the smallest body length of each species $V_0$. $L$ is normalized by the average body width $W$ (SI Table 2). Inset: Comparison of experimental $V(L)$ with the prediction of Model B with a single flagellum $N_f = 1$ (the solid line) (SI Sec.~1B) and simulations of monotrichous bacteria (empty symbols). $f(L)$ is a scaling function, considering the specific geometries of the body and the flagellum of each species (Eq. 17 in SI). The model prediction agrees well with the SBT (red squares) and IB (blue triangles) simulations. ({\it B}) The probability density functions of the length of bacterial body for monotrichous bacteria \textit{V. natriegens} (top) and peritrichous bacteria \textit{E. coli} (bottom). The dashed lines are single-peak fittings as a guide of the eye.}  
\label{fig:fig6}
\end{figure}

\subsection*{Functional benefit of multiflagellarity}

Over billions of years of evolution, peritrichous bacteria have developed a set of extraordinary physical and biochemical mechanisms, regulating the complex dynamics of multiple flagella to achieve swimming in bulk fluids \cite{Wadhwa2022}. Such a daunting task, however, is redundant for monotrichous bacteria with a single rotating flagellum, which accomplish swimming while avoiding the high growth cost for the synthesis of additional flagella. Under strict evolutionary control, multiflagellarity must confer a selective advantage to peritrichous bacteria. One intuitive answer would be that, with more flagella and therefore stronger propulsion, bacteria swim faster. Thus, it came as a surprise when the seminal experiments by Darnton and co-workers suggested that additional flagella barely increase the swimming speed of bacteria \cite{Darnton2007}. These authors then pose an interesting question: ``If they do not allow the cell to swim faster, why does a cell have multiple flagella?'' 

Many answers to the question have been proposed over the last two decades including the role of multiflagellarity in increasing tumbling efficiency \cite{Darnton2007}, enhancing swimming stability \cite{Nguyen2018}, improving search and exploring ability \cite{Najafi2018}, and promoting swarming \cite{Chilcott2000,Ping2010} and biofilm formation \cite{Haiko2013} on solid substrates. We provide here an alternative answer based on the prominent function of flagella in swimming: growing multiple flagella allow bacteria of different sizes to swim at the same speed. As the size variation is unavoidable, if bacteria of larger sizes had lower motility and therefore weaker chemotactic ability \cite{Son2016}, then a colony of bacteria migrating collectively in chemotactic response to an attractant would disintegrate over time and be sorted out spatially according to bacterial sizes. Such ``phenotypic sorting'' decreases the phenotypic heterogeneity of the population, which in turn reduces the overall fitness of the population \cite{Hibbing2010,Waite2018,Ackermann2015}. The size-independent swimming speed enabled by multiflagellarity allows a bacterial colony to migrate collectively while preserving its phenotypic diversity in bacterial sizes.

\subsection*{Monotrichous versus peritrichous bacteria}

The functional benefit of multiflagellarity discussed above has important implications on the swimming behavior and the morphology of monotrichous bacteria. First, the essential role of multiflagellarity in maintaining a constant swimming speed implies that the swimming speed of monotrichous bacteria should decrease with body size. To verify the prediction, we measure the size-speed relation for three species of monotrichous bacteria, \textit{Vibrio alginolyticus}, \textit{Vibrio natriegens} (\textit{V. natriegens}), and \textit{Rhodobacter sphaeroides} (Materials and Methods). In contrast to the constant swimming speed of \textit{E. coli}, all monotrichous species show negative correlations between $V$ and $L$ (Fig.~\ref{fig:fig6}{\it A}), qualitatively agreeing with our model prediction (Fig.~\ref{fig:fig6}{\it A} inset). 

Second, as the decrease of the swimming speed limits the chemotactic ability of monotrichous bacteria of large sizes \cite{Son2016}, the size of monotrichous bacteria should be more tightly regulated with a narrower distribution than the size of peritrichous bacteria. Figure~\ref{fig:fig6}{\it B} compares the probability density function of the size of \textit{V. natriegens} and that of \textit{E. coli}. Indeed, the normalized size variation of \textit{V. natriegens}, defined as the full width at half maximum of the distribution divided by the most probable size of bacteria, is $28\%$, about half of that of \textit{E. coli} at $53\%$. Moreover, the probability to find a \textit{V. natriegens} bacterium larger than its most probable size is $55\%$. This probability is significantly higher at $77\%$ for \textit{E. coli} due to the highly asymmetric shape of their size distribution. Thus, it is less likely to find \textit{V. natriegens} of large sizes compared with \textit{E. coli}, consistent with our prediction based on the trend of the swimming speed. Future studies on more bacterial species in each category are certainly needed to fully verify our hypothesis on the effect of multiflagellarity on the size variation of bacteria.

\medskip

\matmethods{

\subsection*{Bacterial culturing and optical microscopy}

Wild-type \textit{E. coli} strain BW25113 was fluorescently tagged by the insertion of PKK PdnaA-GFP plasmids. A small amount of bacterial stock was inoculated in 2 ml Terrific Broth (TB) [tryptone $1.2\%$ (w/v), yeast extract $2.4\%$ (w/v) and glycerol $0.4\%$ (w/v)] supplanted with $0.1\%$ (v/v) of 100 mg/l ampicillin as an antibiotic. This bacterial solution was then incubated in an orbital shaker at 37 $^\circ$C for 12-16 hours (the late exponential phase). The bacterial culture was further reinoculated and diluted 1:100 with fresh TB and grown in the exponential phase for 6.5 hours in an orbital shaker at 30 $^\circ$C. We harvested motile cells using gentle centrifugation (800$g$, 5 minutes), discarding the supernatant and resuspending the cells in motility buffer [0.01 M potassium phosphate, 0.067 M NaCl, $10^{-4}$ M EDTA, pH 7.0]. The suspension was then washed twice and adjusted to the desired concentration by adding motility buffer. The density of bacteria in motility buffer was measured using a biophotometer via optical density at 600 nm (OD$_{600}$) and controlled at $0.6n_0$, where $n_0 = 8 \times 10^8$ cells/ml is the cell concentration at OD$_{600} = 1.0$. The low density ensured that bacteria do not show collective swimming in our study \cite{Peng2021,Liu2021}. The motile cells were transferred to closed PDMS microchannels of height 150 $\mu$m, where they maintained robust motility for 3 hours at room temperature of 22 $^\circ$C. The measurements were performed at the room temperature within 15 minutes of injecting the cells into the microchannels to ensure constant motility over time. The imaging of bacterial motions was conducted 25 to 30 $\mu$m above the coverslip to avoid potential surface influence using a Nikon Plan APO $60\times$ oil objective (NA = 1.4) on an inverted confocal microscope (Nikon Ti-Eclipse) with a 488 nm laser.  Images were acquired by an Andor zyla sCMOS camera at 30 frames per second.

To image the conformation of bacterial flagella, we cultured \textit{E. coli} strains AW HCB 1, AW HCB 1707 and AW HCB 1732 in Bacto-tryptone broth [10 g Bacto-tryptone, 5 g NaCl, 1 L distilled H$_2$O] that contains minimal free –NH$_2$ groups in medium \cite{Turner2018}. The culturing procedure was the same as that for the wild-type strain. Harvested bacterial cells were washed free of the growth medium by centrifugation (800$g$, 5 minutes) thrice using distilled water. The resulting pellet was resuspended in 500 $\mu$L of distilled water and 25 $\mu$L of 1 M sodium bicarbonate (NaHCO$_3$) was added to raise its pH and allow the optimal coupling of the dye to free –NH$_2$ groups on the bacteria. 1 mg of Alexa Fluor (488 or 532 nm) carboxylic acid succinimidyl esters dye was dissolved in 100 $\mu$L anhydrous DMSO. 25 $\mu$L of this dye was added per 500 $\mu$L of cell suspension. The cell suspension was then stirred on a gentle gyro shaker for 60 minutes in the dark. After 60 minutes, the excess dye was washed out to minimize background fluorescence by centrifugation thrice using distilled water. The flagella were imaged using the same Nikon Plan APO $60\times$ oil objective on a Nikon Ti-Eclipse inverted microscope equipped with a 100W super high-pressure mercury lamp. Images obtained were further processed in ImageJ to enhance contrast for analysis. From the images, we manually measured the axial length $\Lambda$ and the pitch $p$ of the helical flagellar bundle of bacteria (Figs.~S6{\it B} and {\it C}).  

\subsection*{Cell tracking and analysis}

Image stacks from confocal microscopy were processed using the Python library skimage. A GFP-expressing bacterial cell body appears as a bright ellipsoid against a dark background (Fig.~\ref{fig:fig1}{\it A} inset). The video frames were binarized using local intensity thresholding following the Niblack algorithm \cite{Niblack1986}, and the position of the centroid, the orientation, major axis, and minor axis of the cell bodies of all the bacteria were recorded. A standard particle tracking algorithm was then used to link the positions of bacteria across frames, which yielded the full swimming trajectory of individual bacteria. Over the course of a trajectory, each individual bacterium showed variations of their projected length appearing in the video frames. The maximum of the cell length tracked over time was taken as the true length of the bacterium. To reduce noise in tracking, trajectories less than 0.3 seconds were discarded. Since the scanning time of confocal microscopy over the size of a single bacterium was $\sim 0.1$ ms in our experiments, the errors on bacterial length caused by the motion blur of bacterial swimming was neglectable. The average swimming speed of a bacterium over its whole trajectory was obtained, which was then correlated with the length of the bacterium (Fig.~\ref{fig:fig1}{\it B}). The depth of field of the imaging plane of our confocal microscope was about 1 $\mu$m. Thus, as the tumbling of a bacterium typically resulted in the bacterium moving out of the shallow imaging plane, our method dominantly tracked bacterial motions in their run phase. 

To obtain the swimming speed, $V$, we calculated the central finite difference between the positions of bacteria. To measure bacterial wobbling angle $\theta$, we first obtained the temporal evolution of the angle difference between the orientation of bacterial body and the mean direction of swimming given by the unit tangential vector along the bacterial trajectory, $\delta(t)$. The amplitude of the oscillation of $\delta(t)$ was then averaged over the trajectory of the bacterium under consideration, which gave the wobbling angle of the bacterium $\theta$. The correlation between $\theta$ and the length of bacteria $L$ was analyzed and shown in terms of the distribution of wobbling angles at different $L$ in Fig.~S5. The procedures for measuring $V$ and $\theta$ are the same as those used in our previous study \cite{Kamdar2022}. 

\subsection*{Transmission electron microscopy (TEM)}

Transmission electron microscopy (TEM) was used to image the number of flagella of \textit{E.~coli} over a sample size of more than 60 bacteria. Samples for TEM studies were prepared by drop casting bacterial suspensions of concentration $0.1n_0$ onto a lacey carbon film coated copper TEM grid (Pacific GridTech). Negative charge was added to the film surface in a Pelco glow discharger prior to sample deposition to improve yield of bacteria on the imaging area of the film. The grids were air dried for a few minutes prior to the TEM. For visualization of flagella, high-angle annular dark-field (HAADF)-scanning (S)TEM images were obtained on an aberration-corrected FEI Titan 60-300 (S)TEM microscope that is equipped with a CEOS DCOR probe corrector. The microscope was operated at 200 keV with a beam current of 80 pA. The STEM convergence angle used was 25.5 mrad with HAADF detector inner and outer collection angles of 50 and 200 mrad, respectively. Flagellar number was counted manually through the TEM images. 

\subsection*{Monotrichous bacteria}

We cultured \textit{Vibrio natriegens}, a monotrichous marine bacterium, in Lysogeny Broth (LB) supplemented with v2 salts [204 mM sodium chloride, 4.2 mM potassium chloride, and 23.14 mM magnesium chloride]. The suspension in the growth medium was directly observed using the inverted Nikon microscope in the bright-field mode discussed above. The cell size and swimming speed of the bacteria were then tracked manually. The swimming speed of \textit{Rhodobacter sphaeroides}, a monotrichous fresh-water bacterium, was measured from videos posted on the website of the Berg lab \cite{BergWeb}. Prof. Katja Taute measured and kindly provided us the swimming speed of \textit{Vibrio alginolyticus}---another monotrichous marine bacterium---as a function of cell lengths \cite{Taute2015}.

}

\showmatmethods{} 

\acknow{We thank the late Professor Howard Berg for providing us with \textit{E. coli} strains and for fruitful discussion and suggestions. We also thank Katja Taute for helping with the measurements of monotrichous bacteria, Linda Turner for guiding us on fluorescent labelling, Xiao-lun Wu for insightful comments on the trend of $\omega_f$ and Xinliang Xu for the suggestion on the drag coefficients of bacterial body. This research was supported by the U.S. National Science Foundation (NSF) (CBET-2028652) and by the IPRIME program, University of Minnesota. S.K. acknowledges partial funding support from the PPG foundation. M.T.-C. was supported by Clare College, Cambridge (Lynden-Bell Research Fellowship). S.L. was supported by the U.S. NSF (DMS-1853591) and the Charles Phelps Taft Research Center, University of Cincinnati. W.L. was supported by the National Institute for Mathematical Sciences Grant of the Korean government (B22920000). Y.K. was supported by National Research Foundation of Korea Grant (2020R1F1A1A01074981). Part of this work was done at the University Imaging Centers (UIC) and Characterization Facility of University of Minnesota, which receives partial support from the U.S. NSF through the MRSEC (DMR-2011401) and the NNCI (ECCS-2025124) programs. }

\showacknow{} 

\bibsplit[6]

\bibliography{pnas-sample}

\end{document}